# Adaptive-time Synchronization Algorithm for Superlattice Key Distribution

Chengyi Zhang
Tsinghua International School

chriszhang20020905@hotmail.com

Jianguo Xie
Suzhou Institute of Nano-tech and Nano-bionics,
Chinese Academy of Science
xcharles@foxmail.com

**Abstract**

*This paper presents a synchronization algorithm for superlattice key distribution, which is a symmetric encryption solution, by optimizing the Euclidian distance between the two chaotic waveforms generated in the receiver and the sender, respectively. This algorithm based on time synchronization is capable of reconstructing the generated waveforms in the receiver and the sender within the error of 5% to 6% (given the fact that the waveforms were not perfectly congruent when they were originally created).*

## 1. Introduction

The motivation of this paper was to create a synchronization algorithm for the superlattice key distribution system invented by the Chinese Academy of Science (CAS). The superlattice key distribution systems are capable of creating similar keys in the receiver and sender without the limitations of time, distance, and environment [1]. The superlattice key distribution system's keys are not replicable or predictable by a third party unless they gained access to one of these specific pairs of systems. However, superlattice key distribution has a few downsides:

- Superlattice key distribution system creates keys that are not exactly congruent.
- Superlattice key distribution system creates identical keys that are not synchronized in time

Therefore, the superlattice key distribution also needs information reconciliation, which is popularly used in physical key distribution, e.g. quantum key distribution, to make sure that the receiver and sender would have identical keys. The first step in the information reconciliation is to make the keys generated in the receiver and the sender as close as possible within the error of less than 10 %. In order to reach this target, this paper presented a synchronization algorithm and demonstrated experimentally that the difference of the receiver and the sender in superlattice key distribution system can be within 5 % to 8 %, and this synchronization algorithm was ready to practical application for superlattice key distribution.

The current synchronization of superlattice key distribution relies solely on the peak search algorithm, which is an algorithm that synchronizes the keys by identifying the first peaks in the wave forms and matching the wave forms using first peaks. In this algorithm, both the receiver and sender use the peak search algorithm to identify the position of the start of the key. However, the peak search algorithm is not accurate enough because the first peak could have different locations relative to the key. Although the peak search algorithm usually only results in the deviation of only one or two digits, the deviation would accumulate quickly, since 64800 digits of keys was collected for every time. If the hamming distance between the keys exceeds 10 %, they cannot be used for encryption. The synchronization rate of the superlattice key distribution by using the peak search algorithm was not good enough yet.

Given such deficiencies of the peak search algorithm, an extra step needs to be adopted into the synchronization algorithm. Here we proposed a new algorithm by using the first few digits of the key as the synchronization sequence which was sent to the other user for synchronization calibration. Using this synchronization sequence, the sender and the receiver can reach an agreement on the position of the start of the key that followed the end of the synchronization sequence. We called this as the adaptive-time synchronization algorithm.

## 2. Theoretical Analysis

Until now, two synchronization process for the superlattice key distribution system can be implemented by using peak search algorithm as well as adaptive-time synchronization algorithm. Section 2.1 represents the peak search algorithm which was used for synchronization for superlattice key distribution system, and section 2.2 explains the adaptive-time algorithm we proposed for better performance.

### 2.1. Peak Search

Peak Search is a time synchronization algorithm in Superlattice Key Distribution. In a sequence of at least one full-cycle waveform, the front and back of the full-cycle waveform tend to be close to zero, as shown in Fig.1.

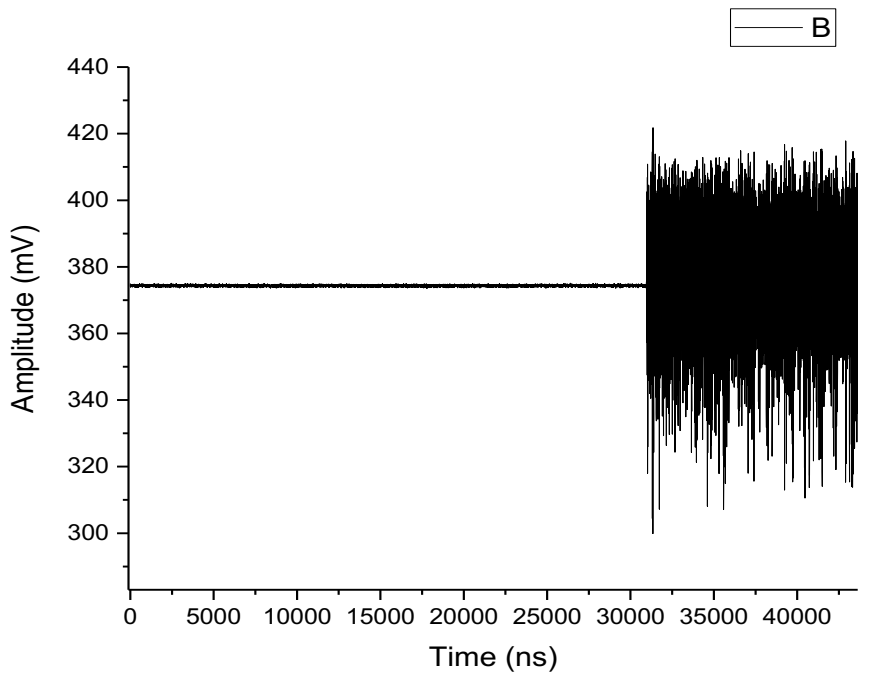

*Figure 1 the waveform of synchronization sequence*

The purpose of the Peak Search algorithm is to find the first peak of waveform B. First, we find the position of the peak by entering values for the parameter $ALIGN_THRESHOLD$ and maximum height of the waveform and scan the waveform starting from zero. However, there are other cases where the starting point of the sequence is in waveform, and also cases where the starting point tends to be zero. As shown above, figure 1 is an example where the starting point tends to be zero and figure 2 is an example where the starting point is in waveform. In the case of figure 1, we designed an algorithm to find the start of the waveform. Find the start of the waveform by setting $ALIGN_WINDOW$ and $SYNC_MARGIN$ to scan the waveform from the starting point. After the start of the waveform is located, we can apply the Peak Search algorithm with the parameters of the location of the waveform to find the first peak in the waveform. In the case of figure 2, we can directly apply the Peak Search algorithm.

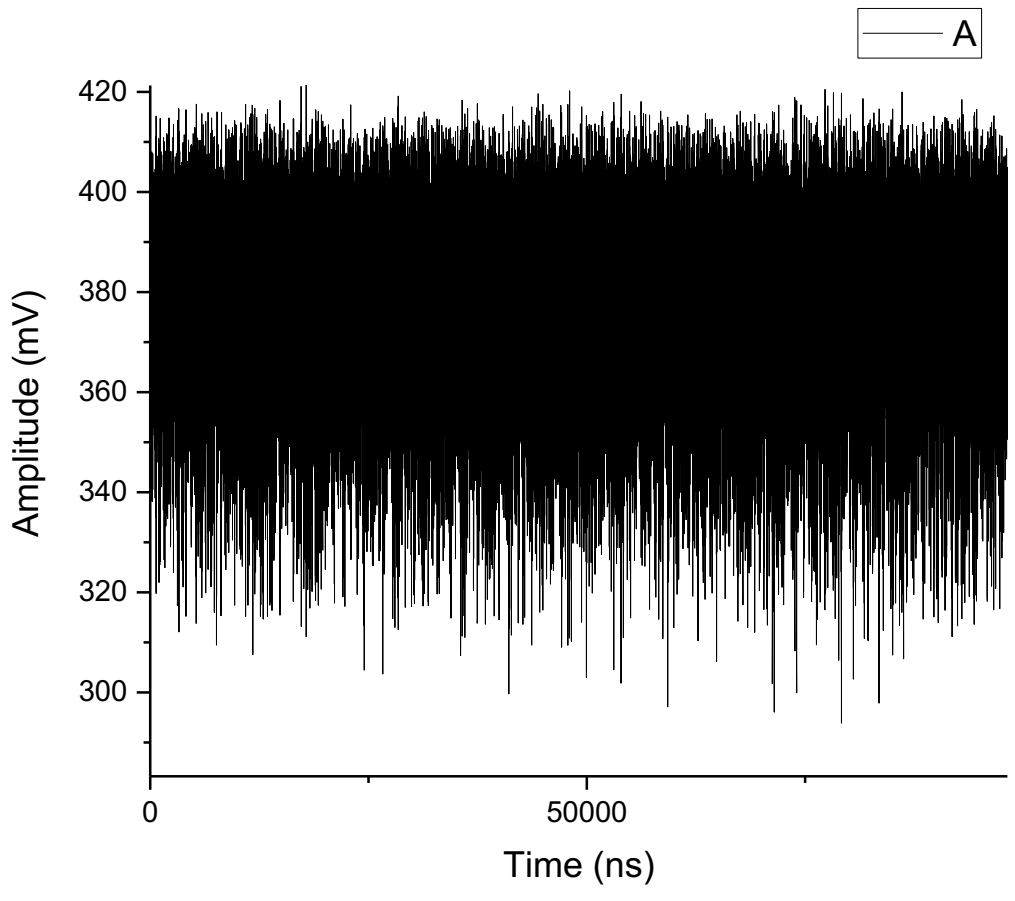

*Figure 2 The original waveform*

**Algorithm 1: Peak Search Algorithm**

1. Initialization. Initialize related parameters such as $ALIGN_THRESHOLD$,$ALIGN_WINDOW$.
2. Determine if the starting point of the sequence is a waveform.
   (a) $if\ sum\big(rawdata(i{:}\,i + ALIGN_WINDOW)\big) > ALIGN_WINDOW * SYNC_MARGIN$
   $i++;$
   (b) $else\ sync_position = i + ALIGN_WINDOW$
3. Find the peak of the waveform
   (a) $for\ i = sync_position : length(rawdata)$
   (b) $if\ rawdata(i) > ALIGN_THRESHOLD * max(rawdata)$
   (c) $if\ rawdata(i) > peakValue$
   (d) $peakValue = rawdata(i),$ $aligndPosition = i$
   (e) $else\ break;$

After the location of the peak is determined, we can then choose a certain length of the waveform following the first peak as the key for encryption. On the receiver side, a similar waveform can be extracted through applying the peak search algorithm on the output from the superlattice key distribution system, which can be used as the key for decryption. It is worth noticing that there is essentially no exchange of information for using the peak search algorithm even though the efficiency and error rate is poor.

## 2.2. Adaptive-Time-Synchronization

The first step of the adaptive time synchronization algorithm is for the sender to select a sequence from its superlattice key distribution system's waveform output and send it to the receiver. For convenience and efficiency, we used the peak search algorithm to locate the start of the waveform and use the first 1200 digits as the synchronization sequence. After the sequence for synchronization is selected from through the peak identification algorithm, we tried to align the synchronization sequence to the corresponding parts in the other system in the superlattice key distribution system's waveform through calculating the Euclidean distance. The algorithm is essentially searching for a segment inside the waveform that has the least Euclidean distance to the synchronization sequence. The algorithm would run from the start of the waveform, cut out a segment that is the same length as the synchronization sequence and calculate the Euclidean distance between the selected segment and the synchronization sequence. Then, a segment is reselected by shifting one digit to the right, and its Euclidean distance is recorded from the synchronization sequence. The previous step is repeated until the last digit can generate a key of 64800 digits too. After the algorithm had gone through all the possibilities, the algorithm will select the starting position which leads to the least hamming distance between the two waveforms as the start of the decryption key for the receiver. The sender will simply use the section of waveform following the synchronization sequence as the encryption key.

**Algorithm 2: Adaptive-Time-Synchronization**

1. Initialize parameters like the synchronization sequence sent from the sender
2. Find the position of the portion of the waveform that is the most similar to the synchronization sequence:

$arr = []$

$for\ i = 1 : length(wave)$

$diff = wave\left(i + length(sync_sequence)\right) - sync_sequence$

$arr.append\ (diff)$

$pos = min(arr)$

The key starts at $pos + length(sync_sequence)$ for the receiver and the key starts right after synchronization sequence for the sender.

3. Experimental Design

In Section 2, we explained how to synchronize the sequence of the keys of the Superlattice Key Distribution System. We proposed an Adaptive-Time-Synchronization algorithm. To further verify the accuracy of our algorithm and its effectiveness in practical applications, we designed the following experiment.

1. Since sequence we send for synchronization may affect the synchronization accuracy, we tested three different driving-sequences or inputs to test the reliability of our algorithm. The three driving-sequences are pseudo-random sequences, periodic sequences, and periodic superposition sequences.
2. Through several experiments and three time-synchronization-sequences, we compare the advantages and disadvantages of Adaptive Time Synchronization algorithm and Peak Search algorithm.

We evaluate the results by the control-variable-method. Hamming distance has been used to characterize the errors between the keys after synchronization in our experiment. The experiment is conducted in the following manner.

1. Generate pseudo-random, periodic, periodic driving-sequences for synchronization, which is then followed by a normal sequence used as an excitation signal.
2. Sequences is collected by Texas Instrument's data acquisition software. Since we were not sure if we would be able to collect a continuous and full-period of waveform by only collecting one period-length of the output, we collected two-period-length of the output which guarantees at least one full period of waveform. The waveform of the three driving sequences are collected. This process is gone through both devices.
3. Synchronize the two waveforms collected on the two machines mentioned above using the Adaptive-Time-Synchronization algorithm. The synchronization results are evaluated by the Hamming distance between the two waveforms after they are synchronized. In this case, a portion of the waveform collected from the sender to the receiver. This process is conducted for all the waveforms collected through the three driving sequences.
4. Synchronize the three waveforms mentioned above using the Peak Search algorithm and evaluate the results by the hamming distance in the waveforms.
5. Comparing the advantages and disadvantages of Adaptive-Time-Synchronization algorithm with Peak Search algorithm by using Hamming distance.

For one driving-sequence in the above experiment, both the sender and the receiver collect 30 output-sequences for synchronization and obtains 900 sets of Hamming distance. We compare and analyze the 900 sets of Hamming distance using statistical methods. For the specific experimental results, please refer to the next Section.

4. Experiment results

For the sake of symmetric encryption scheme, the difference (calculated by the hamming distance) should be less than 8%, which is the criterion we are using for the experiment. The table below (Sheet 1) are the results of the percentages of time in which the respected algorithm failed to synchronize the keys.

*Sheet 1 results of two algorithm*

| | Adaptive time synchronization | Peak Search |
|---|---|---|
| Periodic | 5.78% | 31% |
| Periodic superposition | 6.11% | 15% |
| Pseudo random | 5.89% | 30.90% |

As shown above, the adaptive time synchronization algorithm is a lot more accurate than the peak search algorithm, with only 5.78%, 6.11%, and 5.89% of failure rates, in comparison with 31%, 15% and 30.9% for the peak search algorithm.

In addition to poor accuracy, the peak search algorithm completely failed for the synchronization 30~40 times among 900 times of the experiments as shown below (Sheet 2 and Figure 3).

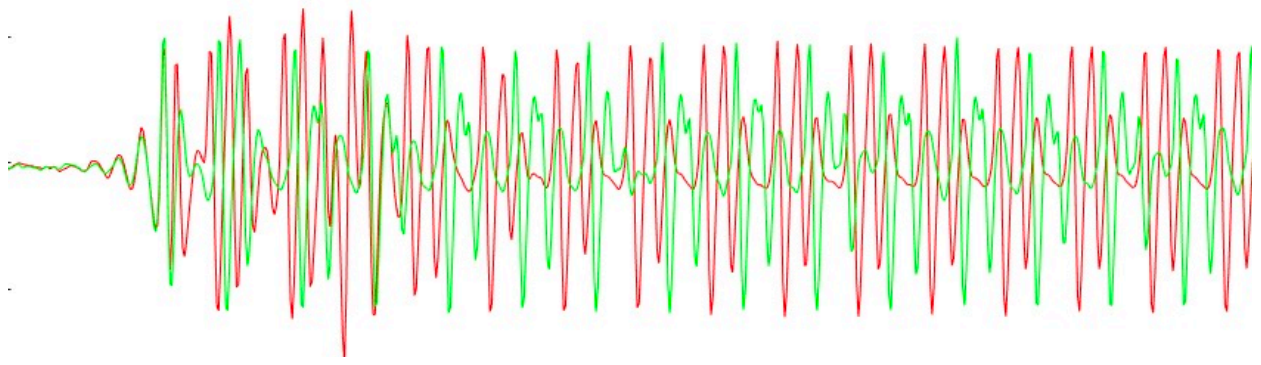

*Figure 3 Result of Peak Search algorithm*

*Sheet 2 Results of Peak Search algorithm*

| | 1 | 2 | 3 | 4 |
|---|---|---|---|---|
| 27 | 28.5119% | 28.33757% | 28.48302% | 28.251157% |

Sheet 2 showed the peak search algorithm's performance when it is dealing with particular sets of periodic superposition sequences, in which the error is as high as 28%. We suspect that the problem with this algorithm is that we need to reset parameters like $ALIGN_WINDOW$ for suspect data, or else the performance will not be optimized. However, as it turns out the problem is associate with superlattice key distribution system itself, as shown below, one of the waveforms was significantly longer than the other. Even though this is already a severe error, such problem is still a reasonable result for the superlattice key distribution system, as it involves complicated physical devices.

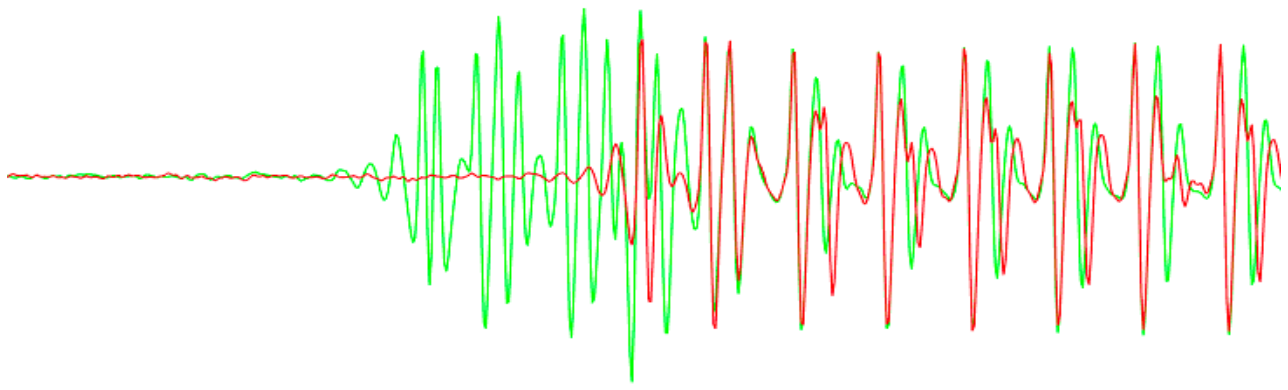

*Figure 4 Result of Adaptive Time Synchronization*

Even in this extreme circumstance, the adaptive time synchronization algorithm still succeeded in synchronizing the two waveforms, with the results shown below (Sheet 3).

Sheet 3 *Results of Adaptive Time Synchronization*

| | 1 | 2 | 3 | 4 |
|---|---|---|---|---|
| 27 | 5.5988% | 2.5768% | 5.8059% | 3.0093% |

This shows that the adaptive synchronization algorithm is a lot more robust comparing to the peek search algorithm when it comes to potential errors in the superlattice key distribution system.

## 5. Conclusion

We presented the adaptive-time synchronization algorithm for superlattice key distribution, which has been demonstrated to be more accurate and robust to extreme circumstances than the previous peak search algorithm. When synchronizing the three sets of data, the adaptive synchronization algorithm maintained a success rate beyond 93%, while the conventional peak search algorithm can have as only a success rate of 70%. Although this algorithm is better in terms of performance, we still have to improve its speed and efficiency, as this algorithm still runs slower than the peak search algorithm. The reason behind the lack of efficiency with the adaptive synchronization algorithm is because it requires intensive calculations on the entire waveform, which is a tedious task comparing to the peak search algorithm, which only requires to go through a small portion of the waveform. Therefore, further improvements of the adaptive synchronization algorithm will be attempts to reduce the number of iterations when determining the starting point of the key.